\documentstyle[paspconf,epsfig]{article} 

\begin{document} 

\title{The mutual
interaction of powerful radio galaxies and their environments}
\author{Christian R. Kaiser\altaffilmark{1}}
\affil{Astrophysics Department, Oxford University, Keble Road, Oxford, OX1 3RH, UK.}
\author{Paul Alexander}
\affil{MRAO, Cavendish Lab., Madingley Road, Cambridge, CB3 0HE, UK.}

\altaffiltext{1}{email: c.kaiser1@physics.oxford.ac.uk}

\begin{abstract}
Using a self-similar model for the expansion of cocoons surrounding the jets
in powerful extragalactic radio sources (type FRII), we investigate the 
influence of the properties of the gas surrounding these objects on their 
evolution. The variation of their radio luminosity as a function of linear 
size for individual sources is determined in dependence of the density 
distribution of the IGM the source is embedded in. Based on these results, 
the cosmological evolution of the FRII population and hence that of the 
environments of these sources can be constrained. The environments are found
to be denser and/or more extended at high redshifts. The bow shock 
surrounding the cocoon of FRII sources compresses and heats the IGM. A 
numerical integration of the hydrodynamic equations governing the gas flow 
between the bow shock and the cocoon boundary is presented. From this we 
determine the appearance of the large scale structure of radio galaxies in 
X-rays and the cooling times for the IGM heated by the passage of the bow
shock. This has important implications for the cosmological evolution of the 
IGM.
\end{abstract}

\section{Introduction}

It has been recognised for many years that powerful extragalactic
radio sources of type FRII are ideal tools for probing the universe at
early epochs. Because of their enormous luminosity in the radio
waveband they can be observed out to very high redshift avoiding the
obscuration by dust which complicates observations at other
frequencies. The properties and appearance of the large scale
structure of FRII sources is influenced by the properties of their
environments. An understanding of how powerful radio galaxies and
their surroundings interact with each other is therefore crucial for
using these objects as local and cosmological probes of the Inter
Galactic Matter (IGM).

\begin{figure}
\centerline{\epsfig{file=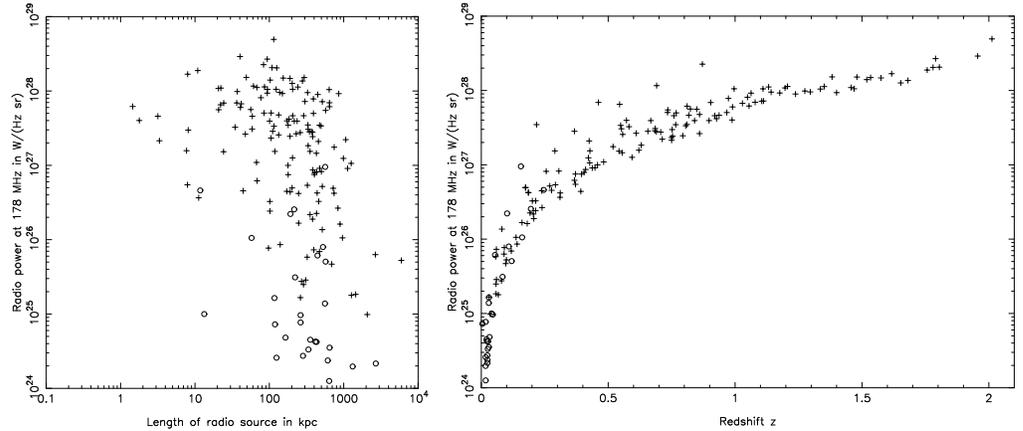, height=13.4cm, angle=270}}
\caption{Source distribution in a flux limited sample of radio galaxies (left) and Malmquist bias (right). All sources are taken from the sample defined by Laing, Riley \& Longair 1983. Crosses: FRII sources, circles: FRI sources. Note that both diagrams have the same scale in radio luminosity which implies that the more luminous sources are located at higher redshift.}
\label{fig:pdzpic}
\end{figure}

Plotting the specific radio luminosity of the large scale structure of
FRII sources (P) against its linear size (D) one obtains a diagram
analogous to the H-R diagram for stars (e.g. Baldwin 1982). However,
the presence of Malmquist bias in any flux limited sample (see Figure
\ref{fig:pdzpic}) means that the source distribution in the P-D
diagram always reflects the combination of the intrinsic evolution of
individual sources and of the cosmological evolution of the FRII
source population as a whole. It is therefore necessary to disentangle
these two effects in order to determine the cosmological evolution of
the radio source environments.

The mechanical energy supplied by an FRII source to its gaseous
environment of order $10^{38}$ W surpasses that of super novae in the
host galaxies of these objects. The bow shock which is propagating
into the IGM in front of the radio cocoon surrounding the jets
compresses and heats the IGM (e.g. Scheuer 1974). This compression
changes the density profile and also the X-ray emission due to thermal
bremsstrahlung of the surrounding gas thereby influencing the
subsequent evolution of this material. In the following we will show
that the resulting X-ray surface brightness profile reflects the
density profile of the unperturbed IGM.

Throughout this paper we will assume $H_o=50$ km s$^{-1}$ Mpc$^{-1}$.

\section{Intrinsic source evolution}
\label{sec:int}

\begin{figure}
\centerline{\epsfig{file=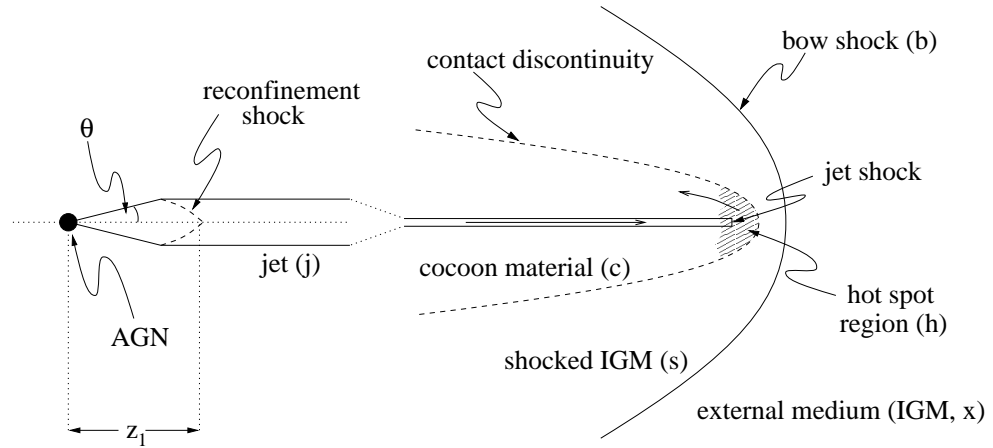, height=13.4cm, angle=270}}
\caption{Basic elements of the large scale structure of an FRII source. Reproduced from Kaiser \& Alexander 1997.}
\label{fig:jetpic}
\end{figure}

The basic elements of the large scale structure of an FRII source are
sketched in Figure \ref{fig:jetpic}. Only one side of a radio source
is shown. The jet is assumed to emerge ballistically from the
AGN. After passing through the reconfinement shock the jet is in
pressure equilibrium with its own cocoon (e.g. Falle 1991, Kaiser \&
Alexander 1997). The jet ends in a strong jet shock and the jet
material subsequently inflates the cocoon. The expansion of the cocoon
is supersonic with respect to the IGM and therefore drives a bow shock
into this material. Because of the high sound speed within the cocoon
the pressure will be uniform in this region except for a small volume
surrounding the jet shock, the hot spot region. The material in this
region has just been injected by the jet into the cocoon and has
therefore had not enough time to `communicate' with the rest of the
cocoon. The higher pressure in the hot spot region explains the
elongated shape of FRII radio galaxies. The density distribution of
the unperturbed IGM is modelled by a simple power law, $\rho _x = \rho
_o (r/a_o)^{-\beta}$, where $\rho _o$ is the density at the core
radius $a_o$ and $r$ is the distance from the centre of the
distribution. Note that this power law is a good approximation to a
King (1972) profile of central density $\rho _o$ in the case of
distances from the centre greater than a few core radii.

This model predicts self-similar growth of the bow shock and the
cocoon. This is supported by observations (e.g. Leahy \& Williams
1984, Black 1992). The derived ages for FRII sources agree well with
observed spectral ages (Kaiser \& Alexander 1997) and the advance
speeds of the radio hot spots are within the limits derived by Scheuer
(1995) from the asymmetries of observed FRII sources.

The radio emission of FRII radio galaxies is created by relativistic
electrons within the cocoon radiating via the synchrotron
mechanism. These electrons are the subject of energy loss processes
which alter the spectrum of the radiation with time. To find the total
radio luminosity of the cocoon the time dependent effects of these
loss mechanisms have to be taken into account. Kaiser, Dennett-Thorpe
\& Alexander (1997) present a model in which the cocoon of FRII
sources is split up into small volumes characterised by the time of
the injection of the relativistic electrons within them into the
cocoon. The effects of adiabatic expansion, synchrotron radiation and
inverse Compton scattering of the Cosmic Microwave Background
Radiation (CMBR) can thereby be traced for these volume elements
independently.

\begin{figure}
\centerline{\epsfig{file=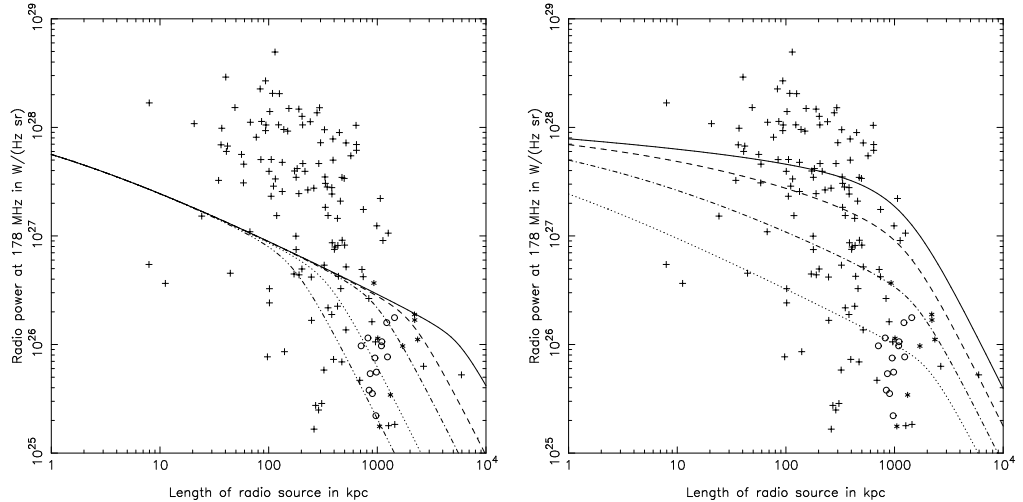, height=13.4cm, angle=270}}
\caption{Evolutionary tracks through the P-D diagram as predicted by the model of Kaiser, Dennett-Thorpe \& Alexander (1997). Left: Effects of varying the source redshift; solid curve: $z=0$, dashed: $z=0.5$, dot-dashed: $z=1$, dotted: $z=2$ and dot-dot-dot-dashed: $z=3$. Right: Varying the external density; the density increases by factors of ten for each curve from bottom to top.}
\label{fig:rhoredcom}
\end{figure}

With this model it is possible to show that the exact details of the
evolution of the magnetic field within the cocoon have no significant
effect on the evolutionary tracks of these sources through the P-D
diagram. Figure \ref{fig:rhoredcom} illustrates the influence of
redshift and the density of the environment on the shape of the
evolutionary tracks. The steepening of the tracks towards larger
linear sizes is caused by the increasing importance of inverse Compton
scattering. Since the energy density of the CMBR is proportional to
$(1+z)^4$, where $z$ is the cosmic redshift, the steepening occurs at
smaller linear sizes for sources at higher redshift. This partly
explains the observation that the mean linear size of sources at
higher redshift is smaller than that of objects at lower redshift (see
Figure \ref{fig:pdzpic}).

The radio luminosity of sources in denser environments is
enhanced. The steepening of the tracks occurs at smaller linear sizes
for these sources because stronger synchrotron losses have already
strongly depleted the number of highly relativistic electrons when
inverse Compton scattering becomes important. The assumed shape of the
density distribution of the IGM implies that central density, $\rho
_o$, and core radius, $a_o$, are not independent parameters. The
evolutionary tracks only depend on a combination, $\rho _o
a_o^{\beta}$, of the two. This means that the effects of a `denser'
environment are indistinguishable from those of a `more extended'
environment and Figure \ref{fig:rhoredcom} may also be interpreted in
this way.

\section{Cosmological evolution}

\subsection{Modelling the source distribution in the P-D plane}

The model for the intrinsic source evolution described in the previous
section depends on source parameters (jet power, aspect ratio of the
cocoon, distribution of energies of the relativistic electrons at
injection, source age, redshift) and parameters describing its
environment (external density parameter $\rho _o a_o^{\beta}$, power
law exponent $\beta$). All of these parameters are either not directly
observable or can only be inferred from observations at comparatively
low redshift. The `birth function' of FRII sources, i.e. the comoving
number density of progenitors of radio galaxies starting to produce
powerful jets as a function of redshift, is of course also unknown. In
the following we will assume reasonable distribution functions of
these source and environment parameters which initially are assumed to
be independent of each other. For the birth function we assume a power
law of the form $(1+z)^n$. For a given cosmology it is then possible
with the help of the model for the intrinsic radio luminosity-linear
size evolution described in the previous section to calculate a
continuous distribution function in the P-D plane. This is then
compared with the observed, binned source distribution of the
complete, flux limited sample of Laing et al. (1983) using a $\chi
^2$-test.

Using this technique we find that the steepening of the evolutionary
tracks of FRII sources due to inverse Compton scattering of the CMBR
at large linear sizes alone is not sufficient to explain the observed
decrease of the median linear size with redshift and/or radio
luminosity. More luminous sources at higher redshift tend to host more
powerful jets and this also implies higher hot spot advance speeds in
these sources. The steepening of their evolutionary tracks therefore
occurs only at larger linear sizes; the opposite of what is observed.

Sources in denser environments will not only be more luminous than
those in more rarefied surroundings but their expansion speed will be
lower as well. The environments of sources at high redshift must have
decoupled from the Hubble flow earlier than those of low redshift
objects. This very simple picture implies that $\rho _o \propto
(1+z)^3$ which leads to a shortening of the mean size of sources at
high redshift. However, we find that this effect is not strong enough.

The single power law assumed for the birth function predicts an
monotonously increasing number of sources with increasing
redshift. For low radio luminosities the flux limit of the comparison
sample means that these sources will not be included in the
sample. For high radio luminosities, however, we find that this birth
function predicts many more radio luminous sources at high redshift
than are observed. This implies at least a flattening, if not a turn
over, of the radio luminosity function at redshifts of around 2 which
is also indicated by observations (e.g. Dunlop \& Peacock 1990).

We find some evidence for a population of giant radio galaxies
distinct from the main population by either their exceptionally high
age and/or very rarefied environments. There are three, possibly four,
sources in the sample of Laing et al. (1983) which have linear sizes
close to or above 1.5 Mpc and radio luminosities below 10$^{26}$ W
Hz$^{-1}$ sr$^{-1}$ which belong to this class. The probability for
finding sources in this region of the P-D diagram is extremely low in
any of the models discussed here and their inclusion in the models by
allowing for extremely high life times ($> 10^9$ years) leads to an
excess of sources with similar sizes but greater radio luminosities
which are not observed (see Figure \ref{fig:pdzpic}).

\begin{figure}
\centerline{\epsfig{file=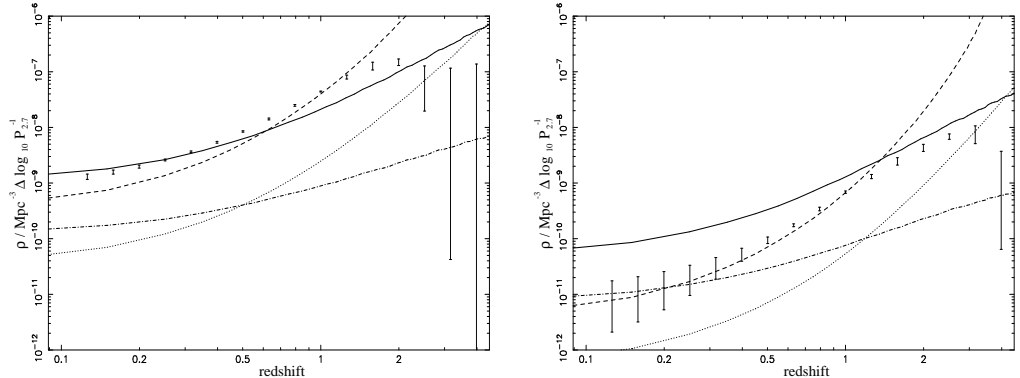, height=13.4cm, angle=270}}
\caption{Radio luminosity function as predicted by the two best fitting models for the cosmological evolution of the FRII source population. Left: 10$^{26}$ W Hz$^{-1}$ sr$^{-1}$ at 2.7 GHz, right: 10$^{27}$ W Hz$^{-1}$ sr$^{-1}$ at 2.7 GHz. Solid lines: $\Omega _o =1$ and model ({\em i}\/) (see text), dashed: $\Omega _o =1$ and model ({\em ii}\/), dot-dashed: $\Omega _o =0$ and model ({\em i}\/) and dotted: $\Omega _o =0$ and model ({\em ii}\/). The error bars show the range of fits to the radio luminosity function using a free modelling approach (Dunlop \& Peacock 1990).}
\label{fig:lumfun}
\end{figure}

The disagreement of the model with the observed source distribution
suggests that it is necessary to relax the assumption that all source
and environment parameters are independent of each other. We find very
good agreement between our models and observations for two models:
({\em i}\/) The maximum life time of a source is proportional to $Q_o
^{-0.5}$, where $Q_o$ is the jet power, or ({\em ii}\/) the density
parameter $\rho _o a_o^{\beta}$ is proportional to $Q_o ^2$. Both
models predict the correct shortening of the median size with redshift
and/or radio luminosity. The radio luminosity function derived from
both models is in good agreement with that derived from observations
out to about $z=1.5$ where our simple birth function starts to
overpredict the number density of high luminosity sources (Dunlop \&
Peacock 1990, see Figure \ref{fig:lumfun}).

\subsection{Interpretation of the cosmological evolution}

The sample of Laing et al. (1983) which we used in the previous
section to constrain the models of the cosmological evolution of the
FRII radio source population includes only the most luminous objects
at any given redshift in the universe. The cosmological evolution of
FRII sources and, possibly, their environments we derive from our
models therefore only applies to these most luminous objects. This
implies that FRII radio galaxies within the model samples defined by
the observational flux limit may have widely different intrinsic
properties at different redshifts. Indeed we find that the median jet
power of sources that should be included in the sample predicted by
the models ({\em i}\/) and ({\em ii}\/) increases strongly with
redshift; $Q_o \propto (1+z)^5$ for model ({\em i}\/) and $Q_o \propto
(1+z)^{5.4}$ for model ({\em ii}\/). Throughout this section it should
be borne in mind that we may compare intrinsically different objects
at different redshifts with each other in this interpretation.

In model ({\em i}\/) the median environment of sources within the
model sample changes only slightly with redshift, $\rho _o a_o^{\beta}
\propto (1+z)^{0.5}$. In model ({\em ii}\/) the environment parameter
is proportional to $(1+z)^{10.8}$. If we set the jet power equal to a
fraction $\epsilon$ of the Eddington luminosity of the black hole in
the centre of the AGN producing the jets in a given source, we find
that $Q_o \propto \epsilon M_{BH}$, where $M_{BH}$ is the mass of the
central black hole. Kormendy \& Richstone (1995) find that the mass of
the central black hole in quiescent galaxies at low redshift is
roughly proportional to the mass of the object it is located in. In
the case of spiral galaxies this is the mass of the central bulge
while for elliptical galaxies, which are the host galaxies of powerful
FRIIs, this is the total mass of the object. The total mass of the
progenitor of the radio source is proportional to the density
parameter of the environment, $\rho _o a_o^{\beta}$. Assuming that the
relation between total mass and mass of the central black hole extends
to higher redshift we find that $\epsilon \propto (1+z)^{4.5}$ in
model ({\em i}\/) and $\epsilon \propto (1+z)^{-4.9}$ in model ({\em
ii}\/). The jet powers of the most luminous FRII sources at $z=1$
inferred from radio observations are comparable to the Eddington
luminosities of black holes of up to $10^9$ solar masses (Rawlings \&
Saunders 1991). In model ({\em i}\/) this implies an efficiency for
the jet production mechanism which is sub-Eddington at low redshift
while in model ({\em ii}\/) $\epsilon$ is increased at lower
redshift. If jets are driven by accretion flows unto black holes, it
is unclear how the jet production mechanism can become super-Eddington
and this may rule out model ({\em ii}\/). However, if $\epsilon$ does
not depend on redshift and, say, $\rho _o$ is constant as well we find
$a_o \propto (1+z)^{10.8/\beta}$ for model ({\em ii}\/). This means
that radio sources should be located in more extended environments at
higher redshift. For a decrease in $\rho _o$ with increasing redshift
this effect becomes even stronger. There is some observational
evidence that the environment of FRII sources may indeed change from
poor groups of galaxies at low redshift to richer environments at
higher redshift (Hill \& Lilly 1991; Best, Longair \& R{\"o}ttgering
1998).

In the scenario of model ({\em ii}\/) the decreasing radio luminosity
of the most luminous radio galaxies with decreasing redshift is caused
by a decrease in the mass of the central black hole within the jet
producing AGN. Since the mass of a super massive black hole will only
increase with cosmic time, there should be plenty of objects in the
local universe with black holes in their centres which are of
comparable or even greater mass than those producing strong radio
sources at high redshift. However, at $z \sim 0$ they only give rise
to weaker jets creating FRI-type sources or they are quiescent
altogether. The most powerful FRII-type objects at low $z$ are always
found in poor groups of galaxies. The faster virialisation of the gas
in large objects like galaxy clusters as opposed to smaller groups
could prevent material from reaching the centre of potential radio
source hosts within these rich environments at low redshift and
thereby depriving the black holes in these objects of fuel
(e.g. Ellingson, Green \& Yee 1991). If this scenario is correct, then
in any attempt to model the cosmological evolution of the radio galaxy
population from hierarchical structure formation it is necessary to
take into account not only the presence of a massive black hole but
also the availability of fuel for the AGN in potential progenitors of
powerful radio galaxies.

To decide which of the two scenarios for the cosmological evolution of
the FRII source population is the more appropriate, particularly at
lower radio luminosities than those covered by the sample of Laing et
al. (1983), it is necessary to use fainter complete samples in the
model comparison. These will be available in the near future and
should help to identify the progenitor population of powerful radio
galaxies and to constrain the cosmological evolution of their
environments.

\begin{figure}
\centerline{\epsfig{file=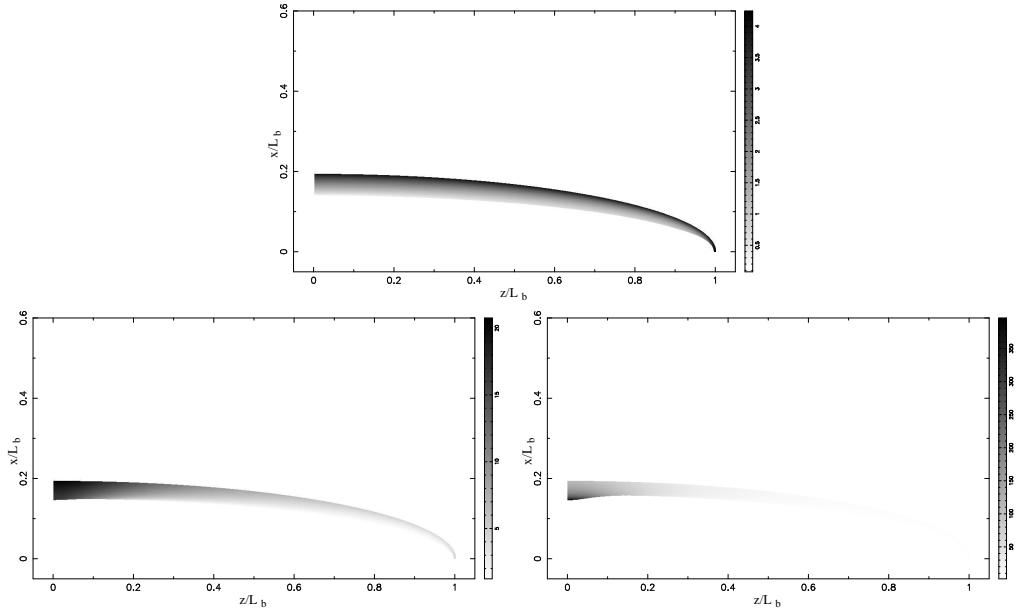, height=13.4cm, angle=270}}
\caption{Density distribution of the gas flow between bow shock and cocoon. Only one half of one cocoon is plotted. The core of the AGN where the jets are produced is located in the lower left corner of each panel at $\rm{z}=0$ and $\rm{x}=0$. The hot spot of the jet is located at $\rm{x}=0$ and z just below 1. The bow shock defines the outer edge of the shaded area while the contact discontinuity delineating the cocoon is the inner edge of the same area. Top panel: uniform external density ($\beta =0$), bottom left: $\beta =1$ and bottom right: $\beta =2$.}
\label{fig:dengray}
\end{figure}

\section{Heating of the IGM}

We have pointed out in section 2 that the expansion of the cocoon of
FRII sources drives a strong bow shock into the surrounding IGM. The
IGM will be compressed and heated by the shock and its properties will
change significantly.

The problem of a strong shock expanding into a gaseous atmosphere was
first solved for the spherical case in a uniform atmosphere by Sedov
(1959). A similar situation but with a constant energy input to a
cavity expanding behind the bow shock was investigated by Dyson, Falle
\& Perry (1980). Both solutions are self-similar and since the
expansion of the cocoon and bow shock of FRII radio sources should be
self-similar as well (Kaiser \& Alexander 1997, Section 2), we expect
that these solutions can be extended to this case. The main difference
is the elongated shape of the cocoons in FRII sources. By assuming
rotational symmetry about the jet axis we can reduce the number of
spatial dimension in the problem by one. Transferring the usual
equations governing the gas flow between bow shock and cocoon to a
self-similarly expanding coordinate system it is possible to transform
these equations to a set of partial differential equations in two
independent (spatial) coordinates. These can then be solved
numerically.

To proceed we have to assume the geometrical shape of the bow
shock. The bow shocks in FRII sources can not be observed
directly. However, we do not expect the layer of shocked gas between
bow shock and cocoon to be very thick and so it seems reasonable to
assume a prolate ellipsoid shape for the bow shock which resembles the
observed shapes of cocoons. At the bow shock surface we assume strong
shock conditions which, together with the assumed power law profile
for the unperturbed density distribution of the IGM, defines the
initial conditions for the integration. The solution is then
propagated numerically inwards from the bow shock and stopped at the
contact discontinuity. The shape of this surface is not known a priori
but can be found using the condition that the gas at the contact
discontinuity is not moving in the direction perpendicular to this
surface.

Figure \ref{fig:dengray} shows the result of the integration for
various values of the exponent of the external density distribution,
$\beta$. The shape of the external density distribution is reflected
in the flow region. For steep external density gradients, $\beta \sim
2$, the density distribution within the flow region is even steeper
than that of the unperturbed gas since in this case the density is
rising in the direction from the bow shock toward the cocoon.

\begin{figure}
\centerline{\epsfig{file=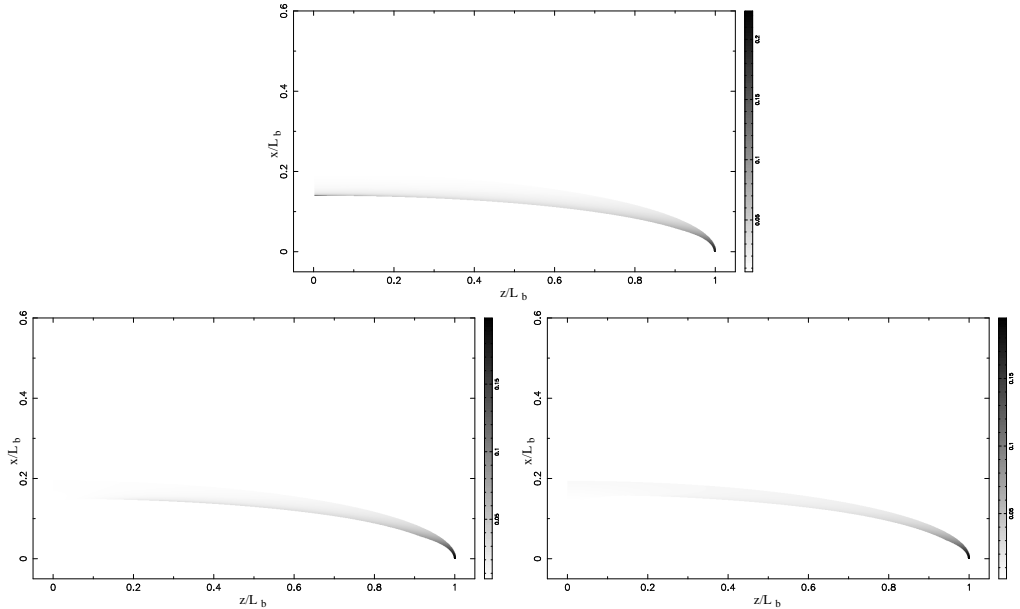, height=13.4cm, angle=270}}
\caption{Temperature distribution in the gas between bow shock and cocoon. The orientation of plots is the same as in Figure \ref{fig:dengray}. Top panel: uniform external density ($\beta =0$), bottom left: $\beta =1$ and bottom right: $\beta =2$.}
\label{fig:temgray}
\end{figure}

We have assumed a power law for the density distribution of the
unperturbed IGM. This can be taken as an approximation to an
isothermal King (1972) model. The isothermal conditions in the
external gas are reflected by the very smooth temperature distribution
within the flow region (Figure \ref{fig:temgray}). Only in front of
the hot spots we find a somewhat higher temperature. 

In section 2 we have shown that the pressure within the cocoon of an
FRII source is constant throughout this region except for a small
region close to the jet shock where the pressure is higher. The
pressure must be matched across the contact discontinuity defining the
boundary of the cocoon. From the analysis presented in this section we
find that the pressure in the flow between bow shock and cocoon does
indeed show a strong decrease in the direction away from the hot spot
region. For a uniform external density the pressure along most of the
length of the cocoon is almost constant. However, for external density
gradients we find the pressure to rise again towards the core of the
radio galaxy. For mild gradients, $\beta \sim 1$, this rise is not
very strong and would be expected because of the predictions by
numerical simulations of backflow of material within the cocoon
(e.g. Norman et al. 1982). In order to avoid gas flow from one side of
the radio source to the other there must be a mild, positive pressure
gradient towards the core to slow down and stop the backflow within
the cocoon. For steeper gradients of the external density this
pressure gradient is also steeper which is inconsistent with the high
sound speed within the cocoon. The analysis presented here will then
apply only as long as the source retains an approximately elliptical
shape. However, it is then unlikely that the expansion is
self-similar. The further evolution of the source depends on the
symmetry of its environment. If the density distribution of the
unperturbed IGM is highly symmetric, the shape of the source may still
be rotationally symmetric about the jet axis but the pressure
equalisation within the cocoon and the region between bow shock and
cocoon will lead to a `pinching' of the cocoon towards the core. The
shape of the bow shock will then not be elliptical. If the IGM is less
symmetrically distributed, significant off axis flow will occur which
is commonly observed in FRII sources.

\begin{figure}  
\centerline{\epsfig{file=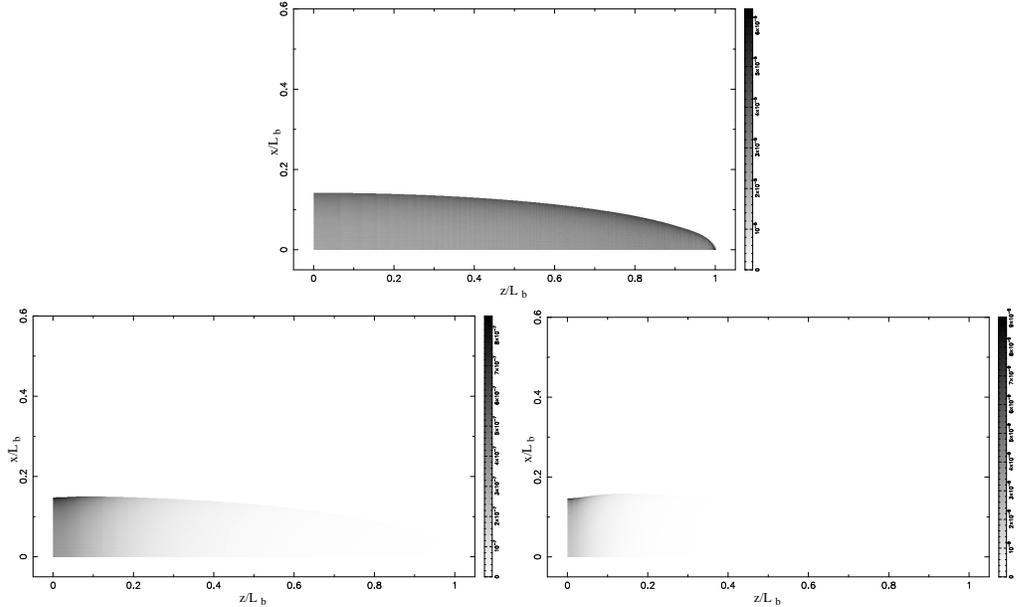, height=13.4cm, angle=270}}
\caption{X-ray surface brightness distribution of the gas between bow shock and cocoon. The orientation of plots is the same as in Figure \ref{fig:dengray}. Top panel: uniform external density ($\beta =0$), bottom left: $\beta =1$ and bottom right: $\beta =2$.}
\label{fig:brigray}
\end{figure}

The X-ray emissivity at frequency $\nu$ in SI units due to thermal
bremsstrahlung of ionised hydrogen is given by (e.g. Shu 1991)

\begin{equation}
\epsilon _{\nu} =7 \times 10^{-51} \, \frac{n_e n_p}{\sqrt{T}} \, e^{-\frac{h \nu}{k T}}\, \frac{\rm{W}}{\rm{m}^3 \, \rm{Hz}},
\label{surbri}
\end{equation}

\noindent where $n_e$ and $n_p$ are the electron and proton number
density respectively and $T$ is the gas temperature. For observations
with a finite bandwidth this is integrated over frequency $\nu$. For
Figure \ref{fig:brigray} we have used the observing band limits of the
High Resolution Imager of the ROSAT satellite (0.1 keV $\rightarrow$ 2
keV). To obtain the X-ray surface brightness for a given source from
this expression we rotate the source about the jet axis and integrate
the emissivity along the line of sight through the source. For
simplicity we assume that the source lies exactly in the plane of the
sky. Because of the exponential expression in equation (\ref{surbri})
the X-ray surface brightness does not scale in a self-similar fashion
and we have to choose parameters like the linear size of the source in
question. In Figure \ref{fig:brigray} we show the surface brightness
for three sources where we have only changed the exponent of the
external density profile between the different panels. Because of the
fixed linear size of the sources plotted here the age of the source in
the constant density profile (top) is about a factor ten higher than
that of the source with $\beta =2$ (bottom right). A comparison with
Figures \ref{fig:dengray} and \ref{fig:temgray} shows that the X-ray
surface brightness is a good tracer of the gas density within the flow
region provided the external gas is isothermal. High resolution, high
sensitivity observations of FRII radio galaxies will therefore enable
us to constrain the properties of the gaseous environments of these
sources at least at low redshift. The cooling times of the shocked IGM
between bow shock and cocoon implied by this calculation are
comparable or exceed the Hubble time for typical jet and environment
parameters. An FRII radio source will therefore influence the
evolution of the gas surrounding it far beyond its own limited life
time.

It is well known that simulations of the formation of galaxy groups
and clusters in the hierarchical structure formation picture predict
gas density distributions with cusps at the centre of the distribution
(e.g. Navarro, Frenk \& White 1995). X-ray observations of the hot gas
in clusters reveal flat density profiles at their centres (e.g. Jones
\& Forman 1984). To reconcile the simulations with the observations
additional sources of heat in the cluster centre are invoked such as
super nova explosions resulting from early star formation in the
cluster progenitors. However, the energy of many super novae is needed
to flatten the density profiles in simulations and these super novae
would produce high values for the metalicity in the gas of galaxy
groups and clusters which is not observed (Ponman in this volume). The
enormous amount of mechanical energy transfered by the jets in FRII
radio sources to the IGM may explain the observed flattening towards
the centre of the density distribution in these objects without
producing any additional metalicity.

\section{Summary}

We have presented models for the intrinsic and cosmological evolution
of radio galaxies of type FRII. The observed decrease of the mean
linear size of these objects with increasing redshift is explained by
two possible scenarios. Either the life time of sources depends on
their jet power or the gas in the environments of sources with more
powerful jets is denser. The second explanation confirms observational
results that FRII radio sources seem to be located in richer
environments at higher redshift. This implies that super massive black
holes in galaxy clusters in the local universe must be starved of fuel
because although they must exist, they do not produce powerful radio
sources.

We have also investigated the properties of the flow of shocked gas of
the IGM between the bow shock and the cocoon of FRII sources. We have
shown that the heating of this material by the bow shock leads to an
enhanced emission of X-rays. The resulting distribution of the X-ray
surface brightness in the flow reflects the density profile of the
material external to the radio source. The expansion of powerful radio
galaxies contributes to the energy of the IGM in clusters and groups
of galaxies and may constitute the additional source of heat necessary
to reconcile numerical simulations of the formation of clusters and
groups with the observed properties of the hot gas in these objects.

\end{document}